\documentclass[aps,pre,twocolumn,array,epsfig,eqsecnum]{revtex4}
\usepackage{amsmath}
\usepackage{amsfonts}
\usepackage[dvips]{graphicx}

\begin{document}   
\setlength{\parindent}{0pt}

\title{Dilapidation of  nonlocal correlations of two qubit states in noisy environment}

\author{K.G. Paulson and S.V.M. Satyanarayana}
\address{
Department of Physics, Pondicherry University, Puducherry 605 014, India}
\date{\today}

\begin{abstract}
Compsite quantum systems exhibit non-local correlations. These counter intuitive correlations form a resource for quantum information processing and quantum computation. In our previous work on two qubit maximally entangled mixed states, we observed that entangled states, states that can be used for quantum teleportaion, states that violate Bell-CHSH inequality and states that do not admit local hidden variable description is the hierarchy in terms of the order of nonlocal correlations. In order to establish this hierarchy, in the present work, we investigate the effect of noise on two quibt states that exhibit higher order nonlocal correlations. We find that dilapidation of nonlocal correlations in the presence of noise follow the same hierarchy, that is, higher order nonlocal correlation disappears for small strength of noise, where as lower order nonlocal correlations survive strong noisy environment. We show the results for decoherence due to amplitude damping channel on various quantum states. However, we observe that same hierarchy is followed by states undergoing decoherence due to phase damping as well as depolarizing channels.

\end{abstract}

\maketitle

\section{Introduction}
Nonlocal correlations exhibited by states of composite quantum systems form and essential resource for different phenomena in quantum information science such as quantum teleportation~\cite{Bennett1993}, quantum cryptography and quantum computation. Nonlocal correlations can be quantified using different measures. Quantum entanglement~\cite{Einstein1935} is the most known measure of nonlocal correlation of a system of particles. Another indicator of nonlocal correlation is the violation of Bell-CHSH inequality ~\cite{Bell1965,CHSH1969}. All pure entangled states violate Bell-CHSH inequality. However, there are mixed entangled state that do not violate Bell CHSH inequality~\cite{Popescu1994}. Werner constructed a mixed state as a convex sum of maximally entangled pure state and maximall mixed separable state~\cite{Werner1989}.

Quantum teleportation refers to transfering an unknown state across two spatially parties which share an entangled bipartite resource. The success of such teleportation is measured in terms of teleportation fidelity. Teleportation fidelity is unity for faithful teleportation and should be more than 2/3 for quantum teleportation. In ~\cite{Horodecki1996} it is shown that states that violate Bell-CHSH inequality can be used as a resource for quantum teleportation. On the other hand, there are states which do not violate Bell-CHSH inequality but are also useful for quantum teleportation. Also there are states that are entangled but not useful for quantum teleportation. Thus, a state that can be used as a resource for quantum teleportation appears to exhibit nonlocal correlations that lie between entanglement and violation of Bell CHSH inequality. Gisin shows~\cite{Gisin1995} if the fidelity of the given state is greater than {\it $F_{lhv}$} (Gisin bound) then the state is non local in the sense, it is incompatible with local hidden variable description. In our previous work~\cite{Paulson 2014} on two qubit maximally entangled mixed states, we observed that entangled states, states that can be used for quantum teleportaion, states that violate Bell-CHSH inequality and states that do not admit local hidden variable description is the hierarchy in which order of nonlocal correlations increases

In this work, we investigate whether the hierarchy in the order of nonlcoal correlations exhibited by two qubit quantum states would preserve itself in the presence of noisy environment. For that we study the decoherence of a pure maximally entangled Bell state, maximally entangled mixed Werner state and a class of states belonging to maximally entangled mixed states due to amplitude damping channel. Let $q$ be the channel parameter denoting the strength of noise. If we begin with a quantum state which exhibits highest order of nonlocal correlation and if $q_G$, $q_B$, $q_T$ and $q_C$ are smallest values of channel parameter for which the teleportation fidelity is less than Gisin bound, Bell CHSH inequality is obeyed, teleportation fidelity is less than 2/3 and the state becomes separable respectively, our main results is $q_G \le q_B \le q_T \le q_C$. Decoherence due to phase damping and depolarizing channels also exhibit the observed hierarchy of nonlocal correlations.

Different measure of nonlocal correlations are defined in section~2. A brief description of noisy quantum channels is given in section~3. Results of dilapidation of nonlocal correlation in maximally entangled pure Bell state and maximally entangled mixed Werner state due to amplitude damping decoherence channel are presented in section~4. Numerical results on the hierarchy exhibited by a collection of maximally entangled mixed states are given section~5. Summary and conclusion are presented in section~6.

\section{Measures of  Non-Locality}
In this section we discuss different measures of nonlocal correlations between two qubits and their quantification methods. We use concurrence ~\cite{Wootters} to calculate the amount of entanglement  of given two qubit state. The analytical expression of concurrence $C$ for density matrix $\rho$ is given as
\begin{equation}\label{c}
C={\rm max} \{ 0,\sqrt{\lambda_{1}}-\sqrt{\lambda_{2}}-\sqrt{\lambda_{3}}-\sqrt{\lambda_{4}}\}
\end{equation}
Where $\lambda^{'s}$ are the eigen values of $\rho\rho\tilde{}$ in the descending order. The spin flipped density matrix $\rho\tilde{}$ is defined as
\begin{equation}
\rho\tilde{}=\sigma_{y}^{A}\otimes\sigma_{y}^{B}\rho^{*}\sigma_{y}^{A}\otimes\sigma_{y}^{B}
\end{equation}
Here $\rho^{*}$ is the complex conjugate of the density matrix $\rho$. C is zero for separable states and $0<C\leq1$ for entangled states.
The optimum teleportation fidelity of a two qubit channel is calculated ~\cite{Horodecki1996} explicitly from
\begin{equation}\label{frho}
F(\rho)= \frac{1}{2}[1+\frac{N(\rho)}{3}]
\end{equation}
where $ N(\rho)=\sum_{i=1}^{3}\sqrt{v_{i}}$ where $\{v_{i}\}$ are eigenvalues of matrix $T^{\dag} T$. T is the correlation matrix and it's elements are determined by $t_{ij}=Tr(\rho\sigma_{i}\otimes\sigma_{j})$. A given state can be used as a resource for quantum teleportation if $F (\rho)>\frac{2}{3}$ or equivalently $N(\rho) > 1$.

A two qubit state obeys Bell CHSH inequality if the Bell parameter is between -2 and 2. The Bell parameter is defined as
\begin{equation}
B_{max}=2\sqrt {M(\rho)}
\end{equation}
where, $M(\rho)=max_{i>j}(v_{i}+v_{j})$.

The measure of nonlocal correlation associated with non-admittance of local hidden variable description is related to optimum teleportation fidelity defined in Eq.(\ref{frho}). If for any state $F(\rho) > F_{lhv}$, such states do not admit local hidden variable description. $F_{lhv}$ is known as Gisin bound and is given by \cite{Gisin1995}
\begin{equation}\label{flhv}
F_{lhv}=\frac{1}{2} + \sqrt{\frac{3}{2}}\frac{{\rm arctan}(\sqrt{2})}{\pi} \approx 0.87
\end{equation}

It was observed \cite{Paulson 2014} that states exhibit a hierarchy in terms of different measures of nonlocal correlations. For example entanglement is necessary for a state to be used as a resource for teleportation but not sufficient. States being teleportation resource is necessary but not sufficient for violation of Bell CHSH inequality. Violation of Bell CHSH inequality is necessary but not sufficient for non admittance of local hidden variable description.

In order to make use of nonlocal correlations of a composite quantum state, two spatially separated parties must share a bipartite state. This can be achieved by two ways. In the first way, one party (Alice) prepares two qubits in a correlated (say entangled) state and sends one quibit to the other party (Bob). The qubit interacts with environmental noise and leads to decoherence of the composite state. This is known as single noisy channel problem. The second way corresponds to a third party prepares two qubits in a correlated state and sends a qubit each to Alice and Bob. In this case both qubits interact with environmental noise, which is called double noisy channel problem. The effect of noisy channel $\Pi$  on a two qubit state $\rho$ in a single channel problem can  be expressed as follows~\cite{M Nielsen},

\begin{equation}\label{ampdamp}
\rho\rightarrow\Pi(\sigma)=\sum_{i}\left(I_2 \otimes M_{i}\right)\rho \left(I_2 \otimes M^{\dag}_{i}\right)
\end{equation}

where $M_i^{'s}$ are the Krauss operators associated with noisy channels. For example, the Krauss operators $M_{0}$ and $M_{1}$ correspond the amplitude damping channel are given as
\begin{equation}\label{krauss}
 M_{0}=\left(
         \begin{array}{cc}
           1 & 0 \\
           0 & \sqrt{1-q} \\
         \end{array}
       \right),
           M_{1}=\left(
               \begin{array}{cc}
                 0 & \sqrt{q} \\
                 0 & 0 \\
               \end{array}
             \right)
               \end{equation}.
where $q$ characterizes the strength of noise and $q=0$ is the ideal noise free condition. We have $0<q\leq 1$.

\section{Non locality of quantum states}
In the present work we consider (a) entangled pure Bell like states (b) maximally entangled mixed Werner states and (c) a class of maximally entangled mixed states proposed by Ishizaka et. al.~\cite{Ishizaka2000}, all with highest order of nonlocal correlations. We subject the above states to decoherence through amplitude damping channel and study dilapidation of different nonlocal correlations.
\subsection{Entangled pure Bell states}
Maximally entangled pure Bell state is defined as
\begin{equation}\label{bell}
\vert\psi\rangle = \frac{1}{\sqrt{2}}\left(\vert01\rangle-\vert10\rangle\right)
\end{equation}

\begin{figure}
  \includegraphics[width=3.5in]{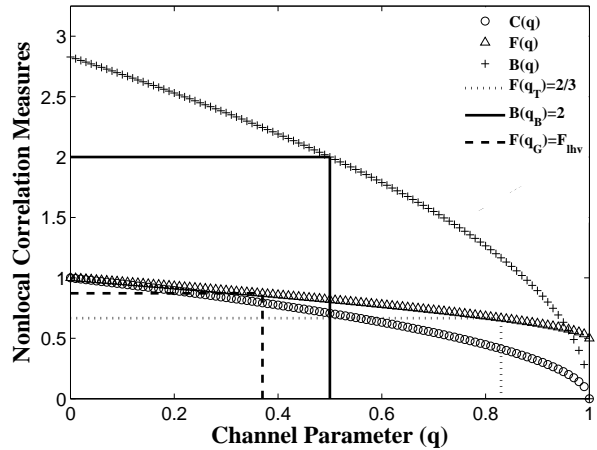}
\caption{Measures of nonlocal correlation as a function of channel parameter $q$ for decoherence of of maximally entangled Bell state through amplitude damping channel.}
\label{fig:1}       
\end{figure}

Decoherence of the state in Eq.(\ref{bell}) through amplitude damping channel is studied using Eq.(\ref{ampdamp}) and Krauss operators defined in Eq.(\ref{krauss}). Analytical expressions for concurrence, optimal teleportation fidelity and Bell parameter are obtained as a function of channel parameter $q$. In fig.\ref{fig:1}, we present concurrence, optimal teleportation fidelity and Bell parameter as a function of strength of the noise $q$. It can be seen from the fig.~\ref{fig:1} that $q_G \le q_B \le q_F \le q_C$. This implies that higher order nonlocal correlations disappears for smaller strength of noise and lower order nonlocal correlations survive high noise strengths. It can be noted in this example that the concurrence is nonzero for all values of $q$ which means that the entanglement survives through the decoherence.
\subsection{Maximally entangled mixed Werner state}
We consider Werner state\cite{Werner1989}, a probabilistic mixture of maximally mixed separable state and maximally entangled pure state, given by
\begin{equation}
|W\rangle=\frac{1-p}{4}I_{4}+p|\psi^{-}\rangle\langle\psi^{-}|,
\end{equation}
where $|\psi^{-}\rangle=\frac{1}{\sqrt2}[|01\rangle-|10\rangle]$ is the maximally entangled Bell singlet state and $p$ is the state parameter $0 \le p \le 1$. As $p$ increases, probability of maximally entangled pure state increases and the state registers higher order nonlocal correlations. Concurrence $C$,
Bell function $B$ and optimal teleportation fidelity $F$ are $\frac{3p-1}{2}$ ,$2\sqrt{2}p$ and $\frac{1+p}{2}$ respectively. From these analytical expressions it is clear that, Werner state is entangled for $p>\frac{1}{3}$ and is useful for quantum  teleportation for the  same values of $p$. The violation Bell-CHSH inequality occurs for $p$ values greater than $\frac{1}{\sqrt{2}}$. The fidelity is more than Gisin's bound for $p>0.74$. A hierarchy in the appearance of nonlocal correlations in Werner state follows $p_{G} \ge p_{B} \ge p_{T} \ge p_{C}$.

We subject Werner state to decoherence analysis through amplitude damping channel. We obtain an exact expression for the concurrence of Werner state as a function of channel parameter and is given by
\begin{equation}\label{wcon}
C(\Pi,W)=p\sqrt{1-q}-\frac{1}{2}\sqrt{(1-p)(1-q)(1-p+q+pq)}
\end{equation}
The teleportation fidelity of Werner state through amplitude damping channel is given by
\begin{equation}\label{wfid}
F(\Pi,W)=\frac{3+(1+2\sqrt{1-q}-q)p}{6}
\end{equation}
Further, the Bell parameter can also be obtained analytically as follows.
\begin{equation}\label{wbell}
B(\Pi,W)=Max{(B_{1},B_{2})}
\end{equation}
where
\begin{equation}\label{b1b2}
B_{1}=2\sqrt{p}\sqrt{1-q} \ \ \ \ \& \ \ \ \ B_{2}=2p\sqrt{2-3q+q^{2}}
\end{equation}

\begin{figure}
\includegraphics[width=3.5in]{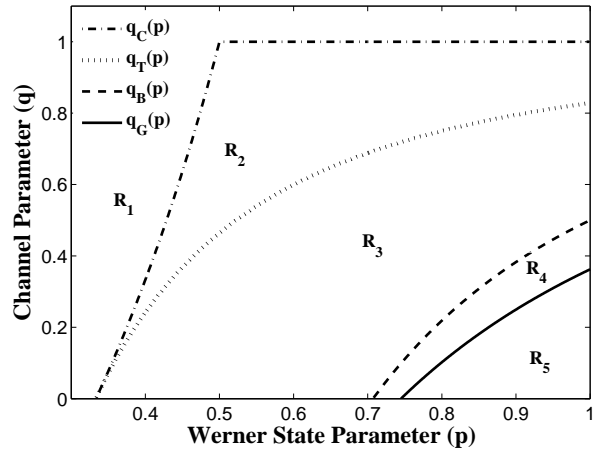}
\caption{Nonlocal correlations of Wener state in the plane of state parameter $p$ and channel parameter $q$}.
\label{fig:2}
\end{figure}

We find $q_C(p)$, $q_F(p)$, $q_B(p)$ and $q_G(p)$ by setting $C=0$, $F=2/3$, $B=2$ and $F=F_{lhv}$ respectively. Curves corresponding to $q_C(p)$, $q_F(p)$, $q_B(p)$ and $q_G(p)$ are shown in fig.\ref{fig:2}. State parameter $p$ - channel parameter $q$ plane can be divided in to regions with respect to order of nonlocal correlations. Region denoted as $R_1$ corresponds to states that are not entangled whose concurrence is less than zero. States in the region $R_2$ are entangled, but cannot be used as resource for quantum teleportation, where optimal teleportation fidelity is less than 2/3. Region $R_3$ consists of states that can be used as resource for quantum teleportation, but do not violate Bell CHSH inequality, whose Bell parameter is less than 2. States in region labeled as $R_4$ violate Bell CHSH inequality, but admit local hidden variable description and states of region $R_5$ are such that their optimal teleportation fidelity is greater than Gisin bound, $F_{lhv}$. It can be clearly seen from fig.\ref{fig:2} that strength of the noise or channel parameter for which dilapidation of different nonlocal correlations follows the same hierarchy as observed in the decoherence of maximally entangled pure Bell state, namely, $q_G \le q_B \le q_F \le q_C$, for all $p$.
\section{Maximally entangled mixed states in noisy environment}
In this section we consider a class of maximally entangled mixed two qubit states defined in~\cite{Ishizaka2000}. These are the states whose entanglement cannot be increased by any unitary transformation. The explicit construction of MEMS proposed by them is given as
\begin{equation}
M=p_{1}|\psi^{-}\rangle\langle\psi^{-}|+p_{2}|00\rangle\langle00|+p_{3}|\psi^{+}\rangle\langle\psi^{+}|\break
+p_{4}|11\rangle\langle11|
\end{equation}
Where $\psi^{\pm}\rangle=\frac{1}{\sqrt{2}}(|01\rangle\pm|10\rangle)$ are bell states, and $|00\rangle$ and $|11\rangle$ are product states orthogonal to $\psi^{\pm}\rangle$. Here $p_{i}^s$ are eigenvalues of M in descending order$(p_{1}\geq p_{2}\geq p_{3}\geq p_{4})$ and $p_{1}+p_{2}+p_{3}+p_{4}=1$. The effect of amplitude damping noise on MEMS is considered and dissipation of non-locality is understood. From fig:3 we understand that the dissipation order of non-locality of MEMS is same as that of above discussed pure and mixed states.
\begin{figure}
\includegraphics[width=3.5in]{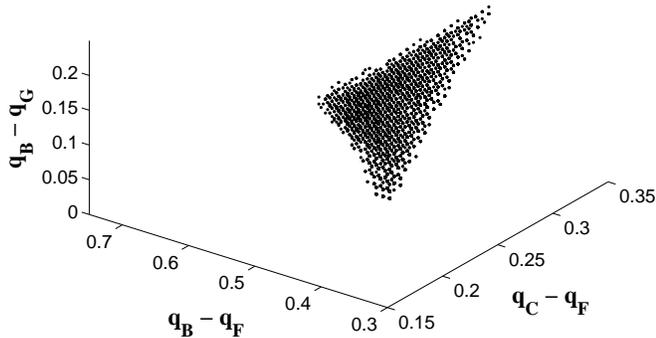}
\caption{ 100000 MEMS proposed by Ishizaka and Hiroshima are generated and $q_{B}-q_{G}$,$q_{B}-q_{F}$ and$q_{F}-q_{C}$ are plotted each other }\label{wn}.
\label{fig:3}
\end{figure}
We randomly generate 100000 two qubit density matrices corresponding to MEMS whose optimal teleportation fidelity is greater than the Gisin bound. We subject each state to decoherence through amplitude damping channel. We collect the values corresponding to $q_G, q_B, q_F$ and $q_C$ for each of the state. In fig.\ref{fig:3}, we plot in three dimensions, the values corresponding to $q_{B}-q_{G}$, $q_{B}-q_{F}$ and $q_{F}-q_{C}$. All points lie in the positive octant of the 3D graph, which means that $q_{B}-q_{G}$,$q_{B}-q_{F}$ and$q_{F}-q_{C}$ are always positive. In the fig.\ref{fig:3}, we have shown only points for 10000 randomly generated states for presentation clarity. This establishes the hierarchy $q_G \le q_B \le q_F \le q_C$ obeyed by this class of MEMS.

\section{Conclusions}
Entanglement is known as nonclassical and nonlocal correlation between two sub systems in the presence superposition. Thus coherence is necessary for nonlocal correlation, but not sufficient. In this paper, we considered different measures of nonlocal correlations for two qubit states and establish a hierarchy among those measures. Entangled states, states that can be used for quantum teleportaion, states that violate Bell-CHSH inequality and states that do not admit local hidden variable description is the hierarchy in which order of nonlocal correlations increases. When states exhibiting highest order nonlocal correlation is subjected to decoherence, it is observed that nonlocal correlations are dilapidates following the same hierarchy, that is higher order nonlocal correlations disappear for small strength of noise, while lower order correlations survive strong noise. We studied decoherence of (a) a maximally entangled pure Bell state (b) a class of Werner states and (c) a class of MEMS through amplitude damping channel. If we denote the smallest values of $q$ for which the optimal telportation fidelity becomes equal to Gisin bound $F_{lhv}$, Bell parameter equal to 2, optimal teleportation fidelity is 2/3 and concurrence is 0 as $q_G, q_B, q_F$ and $q_C$ respectively, we see in all cases that we studied $q_G \le q_B \le q_F \le q_C$. We also investigated and found that states exhibit this hierarchy even if the decohrence is through phase damping channel or depolarizing channel.




\end{document}